\documentclass[prd,showpacs,nofootinbib,preprint,10 pt]{revtex4}
\usepackage{amsmath,graphicx,color,xcolor,epsfig}
\usepackage{epstopdf}
\usepackage{hyperref}

\begin{document}
\begin{titlepage}

\begin{center}
{\bf\Large\boldmath{Two Body Hadronic Decays $\Lambda_{b}(\frac{1}{2}^{+})\rightarrow
B^{\ast}(\frac{3}{2}^{+})+P$ in a quark model}}\\[15mm]
\setlength {\baselineskip}{0.2in}
{\large  Fayyazuddin}\\[5mm]

{\it National Centre for Physics, Quaid-i-Azam University Campus, Islamabad, Pakistan.}\\[5mm]
\end{center}

{\bf Abstract}\\[5mm] 
\setlength{\baselineskip}{0.2in} 
The framework under which decays $\Lambda_{b}(\frac{1}{2}^{+})\rightarrow B(\frac{1}{2}^{+})+M$ are analyzed is not applicable for
the decays $\Lambda_{b}(\frac{1}{2}^{+})\rightarrow B^{\ast}(\frac{3}{2}
^{+})+M$. These decays occur through a baryon pole $\Sigma^{0}_{c}$ which
is generated by the W-exchange diagram in the process $b+u \xrightarrow {W} c+d$. The effective Hamiltonian which arises from the
W-exchange diagram is expressed in the non relativistic limit. Since $%
\Sigma^{0}_{c}$ belongs to representation $6$ of SU(3), it contributes to
two sets of decays: $\Lambda_{b}\rightarrow\Delta^{0}D^{0},\Delta^{-}D^{+},%
\Sigma^{*-}D_s^{+}$ and $\Lambda_{b}\rightarrow\Sigma^{\ast-}_{c}\pi^{+},%
 \Sigma^{\ast 0}_{c}\pi^{0}\;, \Xi^{\ast 0}_{c}K^{0}$. The branching ratios
for these decays are evaluated which can be compared with their experimental
values when the data become available. Other prediction of the model is that
asymmetry parameter $\alpha = 0$, since baryon pole contributes to parity
conserving (p-wave) amplitude and does not contribute to parity violating
(d-wave) amplitude.\end{titlepage}

\section{INTRODUCTION}

 In the standard model two body hadronic decays of heavy flavor
mesons and baryons are analyzed in terms of effective Lagrangian or
Hamiltonian [1,2,3];
\begin{equation}
H_{eff}\equiv V_{cb}V^{\ast}_{qq^{\prime }}[(C_{1}+\frac{1}{3}C_{2})(\bar{q}%
^{\prime\; \beta} q_{\beta})_{V-A}(\bar{c}^{\alpha}b_{\alpha})_{V-A}+(C_{2}+\frac{1}{3}%
C_{1})(\bar{c}^{\beta}q_{\beta})_{V-A}(\bar{q}^{\prime {\alpha}}b_{\alpha})_{V-A}]\label{1}
\end{equation}
where $q=c,q^{\prime }=s$ or $q=u,q^{\prime }=d$. Above Hamiltonian corresponds to decays which are not Cabbibo suppressed. 

The effective Hamiltonian arises from the transition $b\rightarrow c+s+\bar{c}$ or $b\rightarrow c+d+\bar{u}$. The short distance QCD
effects are incorporated in the Wilson coefficient $C_{1}$ and $C_{2}$. In
the factorization ansatz long distance strong interaction effects are
shifted to the evaluation of the baryon form factors in some model.\\
Finally, effective Hamiltonian is written in the form, 
\begin{equation}
H_{eff}=V_{cb}V^{\ast}_{cs}[a_{1}(\bar{s}c)_{V-A}(\bar{c}b)_{V-A}+a_{2}(\bar{%
c}c)_{V-A}(\bar{s}b)_{V-A}]\label{2}
\end{equation}
\begin{equation}
H_{eff} = V_{cb}V^{\ast}_{ud}[a_{1}(\bar{d}u)_{V-A}(\bar{c}b)_{V-A}+a_{2}(\bar{c}u)_{V-A}(\bar{d}b)_{V-A}]\label{3}
\end{equation}
where 
\begin{equation}
a_{1}=C_{1}+\frac{1}{3}C_{2}\; : \; \text{ Tree diagram}\label{4}
\end{equation}
\begin{equation}
a_{2}=C_{2}+\frac{1}{3}C_{1}\; : \; \text{color suppressed tree diagram}\label{5}
\end{equation}
In ref. [4], these decays were analyzed. The form factors which are
functions of $s=q^{2}=(p-p^{\prime})^2$, where $p=p^{\prime }+q$ were
evaluated at the $s=(\text{mass of}\;  M)^{2}$ in quark model, using heavy
quark spin symmetry. In particular using $a_{2}\approx0.10,$ for $%
C_{1}(m_b)\approx1.121$ and $C_{2}(m_b)\approx-0.275$ [3] and recent
experimental values of other parameters, one finds $BR
(\Lambda_{b}\rightarrow\Lambda J/\psi)\simeq1.21\times10^{-4}$ and asymmetry 
$\alpha\simeq-0.18$ to be compared with the experimental values [3]: $%
((6.3\pm0.8)\times10^{-4})$, $\alpha=0.05\pm0.18$. However, replacing $C_2+\frac{1}{3}C_1$, with more general value $C_2+\zeta C_1$ and taking $\zeta = 0.46$ we get $BR = 6.0 \times 10^{-4}$ [4].

It is clear that $(\bar{c}b)_{V-A}$ and $(\bar{s}b)_{V-A}$ in Eq. (2) belong to singlet and triplet representation of $SU(3)$, respectively. Thus in the factorization ansatz, only possible decay modes for $B_{b}$ belonging to representation $\bar{3}$ are $\Lambda_{b} \to \Lambda_{c}^{+}D_{s}^{-}$, $\Xi_{b}^{0}\to \Xi_{c}^{+}D_{s}^{-}$, $\Xi_{b}^{-}\to \Xi_{c}^{0}D_{s}^{-}$ for the first term. For the second term, since $\bar{3}\times3 = 8 + 1$, possible decay modes are $\Lambda_{b} \to \Lambda J/\psi$, $\Xi_{b}^{0}\to \Xi^{0} J/\psi$, $\Xi_{b}^{-}\to \Xi^{-} J/\psi$, where $\Lambda$, $\Xi^{0}$ and $\Xi^{-}$ are members of octet representation of $SU(3)$. 

Hence for the decays $B_{b}(\frac{1}{2}^{+}) \to B^*_{b}(\frac{3}{2}^{+}) P$, where $B_{b}$ belong to representation $\bar{3}$ of $SU(3)$, above framework is not applicable as $ B^*_{b}(\frac{3}{2}^{+})$ either belong to decuplet or sextet representation of $SU(3)$. In this paper, the framework needed for the above decays is formulated.

\section{Effective Hamiltonian for decays $%
\Lambda_{b}(p)\rightarrow B^{\ast}(p')+P(q)$}
For the decays of type $\Lambda _{b}(p)\rightarrow B^{\ast
}(p^{\prime })+P(q)$, the Lorentz structure of $T-$matrix is given by [5]
\begin{equation}
T=\frac{1}{(2\pi )^{\frac{9}{2}}}\sqrt{\frac{mm^{\ast }}{2p_{0}p_{0}^{\prime
}q_{0}}}\frac{q^{\lambda }}{f_{P}}\bar{u}_{\lambda }(p^{\prime })[C-D\gamma
_{5}]u(p)\label{6}
\end{equation}%
where $u_{\lambda }$ is the Raita-Schwinger spinor, $u$ is the Dirac
spinor, and $q = p - p'$. It is clear that $C$ is the parity conserving ($p-$
wave) amplitude and $D$ is the parity-violating ($d-$ wave) amplitude. $f_{P}$ is
the pseudo scalar meson decay constant, which is introduced here to make the
amplitudes $C$ and $D$ dimensionless. For the Rarita-Schwinger spinor $u_{\lambda}(p^{\prime})$
\begin{eqnarray}
\sum_{spin} u_{\lambda}(p^{\prime})\bar{u}_{\mu}(p^{\prime}) &=& \frac{\gamma . p^{\prime}+m^{\prime}}{2m^{\prime}}[\eta_{\lambda \mu}-\frac{1}{3}\gamma_{\lambda}\gamma_{\mu}+\frac{i}{3m^{\prime}}(\gamma_{\lambda}p^{\prime}_{\mu}-\gamma_{\mu}p^{\prime}_{\lambda})-
\frac{2}{3m^{\prime\; 2}}p^{\prime}_{\lambda}p^{\prime}_{\mu}]\notag \\
\overline{\sum_{spin}} u(p)\bar{u}(p) &=& \frac{1}{2}\frac{\gamma . p + m}{2m}\label{7}
\end{eqnarray}
Using above equations and taking traces, we get the decay rate [5]
\begin{equation}
\Gamma =\frac{1}{6\pi }\frac{m|q|^{3}}{m^{\prime 2}f_{P}^{2}}[(E^{\prime
}+m^{\prime })|C|^{2}+(E^{\prime }-m^{\prime })|D|^{2}]\label{8}
\end{equation}%
and for asymmetry $\alpha$ is given by 
\begin{equation}
\alpha=\frac{2|\mathbf{q}|Re CD^{\ast }}{(E^{\prime }+m^{\prime })|C|^{2}+(E^{\prime
}-m^{\prime })|D|^{2}}\label{9}
\end{equation}%
In order to determine the amplitudes $C$ and $D$, one needs basic $H_{eff}$. In the non-leptonic decays of hyperons, there is important
contribution viz the baryon-pole contribution (Born term) to the parity
conserving (p-wave) decay amplitude for which $W$-exchange is relevant [6]. Such a  contribution for $\Lambda_{b}$
decays arises from the W-exchange $b+u \xrightarrow {W} c+d$. 

The effective Hamiltonian for $W-$exchange diagram $b+u \rightarrow c+ d$, is given by
\begin{equation}
H_{eff} = V_{cb} V_{ud}^{*}[(C_1+\frac{1}{3}C_2)(\bar{d}^{\beta}\gamma^{\mu}(1-\gamma^5)u_{\beta})(\bar{c}^{\alpha}\gamma_{\mu}(1-\gamma^{5})b_{\alpha})+(C_2+\frac{1}{3}C_1)(\bar{c}^{\alpha}\gamma^{\mu}(1-\gamma^{5})u_{d})(\bar{d}^{\beta}\gamma_{\mu}(1-\gamma^{5})b_{\beta})] \label{10}
\end{equation}
after taking into account the QCD correction. For the case considered in this paper the first term is relevant. Corresponding to first term, the $M-$ matrix for the $W-$exchange diagram is given by
\begin{equation}
M \approx V_{cb} V_{ud}^{*}\frac{G_{F}}{\sqrt{2}}(C_1+\frac{1}{3}C_2)[\bar{u}(p^{\prime}_{i})\gamma^{\mu}(1-\gamma_5)\alpha_{i}^{+}u(p_{i})][[\bar{u}(p^{\prime}_{j})\gamma_{\mu}(1-\gamma_5)\gamma_{j}^{+}u(p_{j})] \label{11}
\end{equation}
where $q = p_{i}-p_{i}^\prime =  p_{j}^\prime-p_{j}$ and $q^2 < < m^2_{W}$.
In the Pauli representation of $\gamma$ matrices, the four component Dirac Spinor can be written as $u = \begin{pmatrix} u_{A} \\ u_{B} \end{pmatrix}$, where each $u_{A}$ and $u_{B}$ has two components. In the non-relativistic limit $u_{B}$ is of order $v/c$ compared to $u_{A}$. Thus only the bilinears
\begin{eqnarray*}
\bar{u}\gamma^{0}u\approx u^{\dagger}_{A}u_{A} + O(v^{2}/c^2) \notag \\
\bar{u}\gamma^{i}\gamma_{5}u \approx u^{\dagger}_{A}\sigma^{i}u_{A} + O(v^{2}/c^2)
\end{eqnarray*}
are large (see for example [7]).
Using above results, after writing $M$ in terms of two component spinors and then taking the Fourier transform, one gets the effective Hamiltonian in the leading non-relativistic limit for the $W-$exchange [5,6]. 
\begin{eqnarray}
H^{pc}_{W}&=&\frac{G_{F}}{\sqrt{2}}V_{cb}V^{*}_{ud}(C_1+\frac{1}{3}C_2)\sum_{i\neq
j}\alpha^{+}_{i}\gamma^{-}_{j}(1-\mathbf{\sigma_{i}.\sigma_{j}})\delta^{3}(x)\notag \\
H^{pv}_W &=& 0\hspace{12cm} \label{12}
\end{eqnarray}
where $\alpha_{i}^{+}$ and $\gamma^{-}_j$ are operators which convert $b-$quark to $c-$quark and $u-$quark to $d-$quark. 
\begin{eqnarray*}
\alpha_{i}^{+}|b\rangle = |c\rangle \notag \\
\gamma_{j}^{-}|u\rangle = |d\rangle.
\end{eqnarray*} 
Following comments are in order.
$H_{W}^{pc}$ in the leading non-relativistic limit was first derived in ref. [6] for the parity conserving non-leptonic decays of baryons. The result obtained insure $\Delta I = 1/2$ rule (or octet dominance) in agreement with experiment. Other results obtained were also in agreement with the experimental values.

In ref. [5], the decays $\Lambda_{c}\to \Delta^{++}K^{-}\; , \Sigma^{*\; 0}\pi^{+}\; , \Xi^{*\; 0}K^{+}$, were analysed in the same framework. The branching ratio: $BR (\Lambda_{c}\to \Delta^{++}K^{-}) = 9.0 \times 10^{-3}$ obtained is in good agreement with the experimental value $(8.6 \pm 1.0)\times 10^{-3}$. For the case, analyzed in this paper, the $W$-exchange diagram $b+u \to c+d$, involves two heavy quarks, hence the leading non-relativistic approximation valid upto $O(v^2/c^2)$ is viable to apply.
We note that our approach has some analogy with that considered in ref. [8]: Finally, QCD correction has also been incorporated, it gives a factor $C_1+\frac{1}{3}C_2)^2 \approx 1.06$

Before we proceed further we note that [9] 
\begin{equation}
|\Lambda^{0}_{b\;, c}\rangle=\frac{1}{\sqrt{2}}|(ud-du)b\;, c>\chi_{M A}\label{13}
\end{equation}
Thus $\Lambda_{b}$ and $\Lambda_{c}$ belong to triplet rep. $\bar{3}$ of
SU(3). The spin ${\frac{1}{2}}^{+}$ and spin ${\frac{3}{2}}^{+}$ baryons
which belong to representation 6 of SU(3), are 
\begin{equation}
S_{ij}(S^{*}_{ij})=\frac{1}{2}|(q_{i}q_{j}+q_{j}q_{i})c>\chi_{M S}(\chi_{S})\label{14}
\end{equation}
where $q = u, d, s$. The spin wave function $\chi_{M A}, \chi_{M S}$ and $\chi_{S}$
are [9]. 
\begin{equation}
\chi_{M A}=\frac{1}{\sqrt{2}}|(\uparrow\downarrow-\downarrow\uparrow)\uparrow>%
\label{15}
\end{equation}
\begin{equation}
\chi_{M S}=\frac{1}{\sqrt{6}}|-(\uparrow\downarrow+\downarrow\uparrow)%
\uparrow+2\uparrow\uparrow\downarrow>\label{16}
\end{equation}
\begin{equation}
\chi_{S}=\frac{1}{\sqrt{3}}|\uparrow\uparrow\downarrow+(\uparrow\downarrow+%
\downarrow\uparrow)\uparrow>\label{17}
\end{equation}
In particular, we note that 
\begin{equation}
S_{22}=\sqrt{2}ddc \chi_{M S}=\sqrt{2}\Sigma^{0}_{c}\label{18}
\end{equation}
It is clear that relevant operators are  
\begin{equation}
\alpha^{+}_{3}\gamma^{-}_{1}(1-\mathbf{\sigma_{1}.\sigma_{3}})+\alpha_{3}^{+}\gamma_2^{-}(1-\sigma_{2}.\sigma_{3}) \label{19}
\end{equation}
Hence we get
\begin{equation}
\sum_{i\neq j}\alpha^{+}_{i}\gamma^{-}_{j}(1-\mathbf{\sigma}_{i}.\mathbf{\sigma}_{j}%
)|\Lambda_{b}>=\sqrt{6}|\Sigma^{0}_{c}>\label{20}
\end{equation}
We note 
\begin{equation*}
\bar{3}\times10=6+24\;, 8\times6=\bar{3}+6+15+24
\end{equation*}
Thus only possible decays through $\Sigma^{0}_{c}$ pole are

Set I (II)
\begin{equation*}
\Lambda_{b}\rightarrow\Sigma^{0}_{c}\rightarrow\Delta^{0}D^{0},%
\Delta^{-}D^{+},\Sigma^{*-}D^{+}_{s}\; (\Sigma^{*-}_{c}\pi^{+},%
\Sigma^{*0}_{c}\pi^{0}\; , \Xi^*_{c}K^{0})
\end{equation*}
Hence in SU(3) limit, the p-wave (parity conserving) amplitude $C$ for the two sets of decays is given by: \newline
\begin{equation}
C=F(C_1+\frac{1}{3}C_2)\frac{<\Sigma^{0}_{c}|H^{pc}_{W}|\Lambda_{b}>}{%
m_{\Lambda_{b}}-m_{\Sigma_{c}}}\label{21}
\end{equation}
where $F$ is $(2,2\sqrt{3},2)g$ and $(\sqrt{2},-\sqrt{2},\sqrt{2})g_c$ for Set I and II, respectively.
The weak matrix elements $<\Sigma^{0}_{c}|H^{pc}_{w}|\Lambda_{b}>$ on using
Eq. (12) and Eq. (20) is given by: \newline
\begin{equation}
<\Sigma^{0}_{c}|H^{pc}_{W}|\Lambda_{b}>=[\frac{G_{F}}{\sqrt{2}}%
V_{cb}V^{*}_{ud}]\sqrt{6}d^{\prime }\label{22}
\end{equation}
where [8], %
\begin{equation}
d^{\prime }=<\psi_{0}|\delta^{3}(x)|\psi_{0}>=\frac{3(m_{\Delta}-m_{\Sigma_{c}})}{8\pi \alpha_{s}}\bar{m}^{2} \label{23}
\end{equation}
for Set I and for Set II: 
\begin{equation}
d^{\prime }=\frac{3(m_{\Sigma^{*}_{c}-m_{\sum_{c}}})}{8\pi\alpha_{s}}%
\bar{m}^{2}_{c} \label{24}
\end{equation}

Since $H_{W}^{p.v.} = 0$, it follows that baryon pole can not generate $d-$wave amplitude, hence $D=0$ and thus asymmetry $\alpha = 0$. This is in accordance with two particle non-leptonic decay of $\Omega^{-}$ for which the experimental value of $\alpha =0$. This is the first prediction of framework used without detailed analysis.
\section{Detailed analysis of $\Lambda_{b}\rightarrow\Delta D
$ and $\Lambda_{b}\rightarrow\Sigma^{*}_{c}\pi$}

We first discuss the decay $\Lambda _{b}\rightarrow \Delta D$ decay. In the rest frame of $\Lambda _{b},|\mathbf{q}|=2.327$
Gev for the final state $\Delta ^{0}D^{0},\Delta ^{-}D^{+}$ and $|\mathbf{q}%
|=2.242$ Gev for $\Sigma^{\ast -}D_{s}^{+}$. Using
experimental values for $m=m_{\Lambda _{b}}$ and $m^{\ast }=m_{\Delta }$ or $%
m_{\Sigma^{\ast }}$ $f_{D}$ = 207 MeV and $%
f_{D_s}$ = 257 MeV [10], we get from Eq. (8) 
\begin{equation}
\Gamma=2.17\times 10^{2}|C|^{2}GeV\label{25}
\end{equation}%
\begin{equation}
=1.13\times 10^{2}|C|^2GeV\label{26}
\end{equation}%
for $\Delta ^{0}D^{0},\Delta ^{-}D^{+}$ and $\Sigma^{\ast -}D_{s}^{+}$ First we note that constant $g$ in Eq. (24) can
be estimated by using PCAC (partial conservation of axial vector current)
and NQM (non-relativistic quark model): 
\begin{equation}
g=g_{\Sigma_{c}^{0}D^{0}\Delta ^{0}}=g_{\Sigma_{c}^{0}D^{+}\Delta ^{-}}=\frac{%
m_{\Sigma_{c}^{0}+m_{\Delta }}}{f_{D}}g_{A}\label{27}
\end{equation}%
Now in NQM [11], $g_{A}=-\frac{2\sqrt{2}}{3\sqrt{3}}$. Thus 
\begin{equation}
g=9.686\label{28}
\end{equation}%
and 
\begin{equation}
g=g_{\Sigma_{c}^{0}D_{s}^{+}\sum^{\ast -}}=\frac{m_{\Sigma_{c}^{0}}+m_{\Sigma^{+}}%
}{f_{D_s}}g_{A}=8.024\label{29}
\end{equation}
In Eq. (24) in order to take into account large momentum transfer in heavy
flavor baryon decays, we take 
\begin{equation}
\bar{m}^2 = \frac{m_{d}^{4}}{(m_{b}+m_{c})^{2}} \approx 3.18 \times 10^{-4}\label{30}
\end{equation}
Using constituent quark masses, $m_{d}\approx 0.334 GeV\;, m_{c}\approx 1.4GeV\;, m_{b}\approx 4.85GeV\;, \alpha_{s}\approx 0.32$ and experimental values
\begin{equation}
m_{\Sigma_c}-m_{\Delta} \approx 1.22 GeV\; , m_{\Sigma_c}-m_{\Sigma^*} \approx 1.07 GeV \label{31}
\end{equation}%
we get from Eqs. (22, 23) and Eqs. (28-30)%
\begin{equation}
C\approx 7.75\times 10^{-10},\text{ \ \ \ }5.62\times 10^{-10}\label{32}
\end{equation}%

Hence from Eqs. (25, 26):%
\begin{equation*}
\Gamma (\Lambda _{b}\longrightarrow \Delta ^{0}D^{0})\approx 1.23\times
10^{-16}GeV
\end{equation*}%
\begin{equation}
\Gamma (\Lambda _{b}\longrightarrow \Delta ^{-}D^{+})=3\Gamma (\Lambda
_{b}\longrightarrow \Delta ^{0}D^{0})\approx 3.69\times 10^{-16}GeV
\label{33}
\end{equation}%
\begin{equation}
\Gamma (\Lambda _{b}\longrightarrow \Sigma^{\ast -}D_{s}^{+})\approx
3.38\times 10^{-17}GeV  \label{34}
\end{equation}%
Using $\tau _{\Lambda _{b}}\approx 1.451\times 10^{-12}s,$
we get
\begin{equation}
BR(\Lambda _{b}\longrightarrow \Delta ^{0}D^{0})\approx 2.70\times 10^{-4},%
\text{ \ \ \ \ }B.R(\Lambda _{b}\longrightarrow \Delta ^{-}D^{+})\approx
8.11\times 10^{-4}  \label{35}
\end{equation}%
\begin{equation}
BR(\Lambda _{b}\longrightarrow \Sigma^{\ast -}D_{s}^{+})\approx
7.04\times 10^{-5}  \label{36}
\end{equation}%
Set II%
\begin{equation*}
\Lambda _{b}\longrightarrow \Sigma_{c}^{\ast -}\pi^{+},\text{ }%
\Sigma_{c}^{\ast 0}\pi ^{0},\text{ }\Xi _{c}^{\ast 0}K^{0}
\end{equation*}%
\begin{equation*}
|\vec{q}|=2.244GeV,\text{ \ \ }2.240GeV
\end{equation*}%
 From Eq. (8), using $f_{\pi}\approx 131 MeV$ and $f_{K}\approx 160 MeV$, we get%
\begin{equation}
\Gamma \left( \Lambda _{b}\longrightarrow \Sigma_{c}^{\ast -}\pi
^{+}\right) \approx 1.90\times 10^{2}|C|^{2}GeV  \label{37}
\end{equation}%
\begin{equation}
\Gamma \left( \Lambda _{b}\longrightarrow \Xi _{c}^{\ast 0}K^{0}\right)
\approx 1.21\times 10^{2}|C|^2GeV  \label{38}
\end{equation}%
Now PCAC gives%
\begin{eqnarray*}
g_{c} &=& g_{\Sigma_{c}^{\ast -}\Sigma_{c}^{0}\pi ^{+}}=\frac{%
m_{\Sigma_{c}^{\ast }}+m_{\Sigma_{c}}}{f_{\pi }}g_{Ac}\approx 25.30 
\notag \\
g_{\Xi _{c}^{\ast 0}\Sigma_{c}^{0}K^{0}} &=&\frac{m_{\Xi
^{0}}+m_{\Sigma_{c}}}{f_{K}}g_{Ac}\approx 21.24
\end{eqnarray*}%
on using NQM value $g_{Ac}=2/3$.
Hence from Eqs. (22, 24), using %
\begin{equation*}
\bar{m}_{c}^{2}=\left( \frac{m_{d}m_{c}}{m_{b}+m_{c}}\right) ^{2}\approx
5.60\times 10^{-3}
\end{equation*}%
and%
\begin{equation}
(m_{\Sigma_{c}^{\ast }}-m_{\Sigma_{c}}) \approx 0.064 GeV \; , 
( \Xi_{c}^{\ast }-m_{\Sigma_{c}})\approx 0.191GeV  \label{39}
\end{equation}%
we get%
\begin{equation}
C\approx 1.31\times 10^{-9}\text{ \ \ and \ \ }C\approx 3.30\times 10^{-9}
\label{40}
\end{equation}%
 for $\Sigma_{c}^{\ast -}\pi ^{+}$ and $\Xi _{c}^{\ast
0}K^{0}.$

Hence for the decay rates and the branching ratios, we get from Eqs.
(37) and  (38)%
\begin{equation}
\Gamma \left( \Lambda _{b}\longrightarrow \Sigma_{c}^{\ast -}\pi
^{+}\right)  = \Gamma \left( \Lambda _{b}\longrightarrow
\Sigma_{c}^{\ast 0}\pi ^{0}\right) \approx 2.92\times 10^{-16}GeV \label{41} 
\end{equation}
\begin{equation}
\Gamma \left( \Lambda _{b}\longrightarrow \Xi_{c}^{\ast 0}K^{0}\right) 
\approx 12.46\times 10^{-16}GeV  \label{42}
\end{equation}
\begin{equation}
BR\left( \Lambda _{b}\longrightarrow \Sigma_{c}^{\ast -}\pi
^{+}\right)  = B.R\left( \Lambda _{b}\longrightarrow
\Sigma_{c}^{\ast 0}\pi ^{0}\right) \approx 6.47\times 10^{-4}
\label{43}
\end{equation}
\begin{equation}
BR\left( \Lambda _{b}\longrightarrow \Xi_{c}^{\ast 0}K^{0}\right) 
\approx 2.74\times 10^{-3}  \label{44}
\end{equation}%
To conclude: No experimental data for the branching ratios of two
set of decay channels:%
\begin{equation*}
\Lambda _{b}\longrightarrow \Delta ^{0}D^{0},\text{ }\Delta ^{-}D^{+},\text{ 
}\Sigma_{c}^{\ast -}D_{s}^{+}
\end{equation*}%
and%
\begin{equation*}
\Lambda _{b}\longrightarrow \Sigma_{c}^{\ast -}\pi ^{+},\text{ }%
\Sigma_{c}^{\ast 0}\pi ^{0},\text{ }\Xi _{c}^{\ast 0}{K}^{0}
\end{equation*}%
are available to test the branching ratios given in Eq. (35, 36) and Eqs.
(43, 44). One notes that relative branching ratios viz%
\begin{equation}
\frac{\Gamma (\Lambda _{b}\longrightarrow \Delta ^{-}D^{+})}{\Gamma (\Lambda
_{b}\longrightarrow \Delta ^{0}D^{0})}\approx 3,\text{ \ \ }\frac{\Gamma
(\Lambda _{b}\longrightarrow \Sigma^{\ast -}D_{s}^{+})}{\Gamma
(\Lambda _{b}\longrightarrow \Delta ^{0}D^{0})}\approx 0.28  \label{45}
\end{equation}%
and%
\begin{equation}
\frac{\Gamma (\Lambda _{b}\longrightarrow \Sigma_{c}^{\ast 0}\pi ^{0})%
}{\Gamma (\Lambda _{b}\longrightarrow \Sigma_{c}^{\ast -}\pi ^{+})}%
\approx 1,\text{ \ \ \ }\frac{\Gamma (\Lambda _{b}\longrightarrow \Xi
_{c}^{\ast 0}K^{0})}{\Gamma (\Lambda _{b}\longrightarrow
\Sigma_{c}^{\ast -}\pi ^{+})}\approx 4.28  \label{46}
\end{equation}%
\newline
are independent of the parameters $\bar{m}^{2},\bar{m}_{c}^{2}$ and the
axial vector coupling constants $g_{A}$ and $g_{Ac}$. Thus Eqs. (45, 46)
together with prediction that asymmetry parameter $\alpha = 0$ will test the
general frame work used in the analysis of decays $
\Lambda _{b}(\frac{1}{2}^{+})\rightarrow B^{\ast }(\frac{3}{2}^{+})+P$.

Finally decays with three particles in the final state through
resonances:%
\begin{eqnarray*}
\Lambda _{b} &\rightarrow &\Delta ^{0}D^{0}\rightarrow p\pi ^{-}D^{0} \\
 &\rightarrow &\Delta ^{-}D^{+}\rightarrow n\pi ^{-}D^{+} \\
 &\rightarrow &\Sigma^{\ast -}D_{s}^{+}\rightarrow
\Lambda \pi ^{-}D_{s}^{+}\\
&\rightarrow&\Sigma ^{0}\pi ^{-}D_{s}^{+}
\end{eqnarray*}%
are of considerable interest. For the decuplet $\Delta ,m_{\Delta }=1232$
MeV, $\Gamma _{\Delta }\approx 117${\large MeV, }$m_{\Sigma^{\ast
}}=1385${\large MeV, SU(3) gives [1]:}%
\begin{eqnarray*}
\Delta ^{0} &\rightarrow &p\pi ^{-}: \sqrt{2}g^{\ast } \\
&\rightarrow &n\pi ^{0}=2g^{\ast } \\
\Delta ^{-} &\rightarrow &n\pi ^{-}: \sqrt{6}g^{\ast } \\
\Sigma_{c}^{\ast -} &\rightarrow &\Sigma^{-}\pi ^{0}: g^{\ast }
\\
&\rightarrow &\Sigma^{0}\pi ^{-}:g^{\ast } \\
&\rightarrow &\Lambda \pi ^{-}:-\sqrt{3}g^{\ast }
\end{eqnarray*}%

Using physical masses and phase space factor:
\begin{eqnarray*}
\Gamma (\Delta ^{0} &\rightarrow &p\pi ^{-})\simeq 39MeV,\text{ \ \ }\Gamma
(\Delta ^{0}\rightarrow n\pi ^{0})\simeq 78MeV \\
\text{\ }\Gamma (\Delta ^{-} &\rightarrow &n\pi ^{-})\simeq 117MeV \\
\Gamma (\Sigma^{\ast -} &\rightarrow &\Lambda \pi ^{-})\simeq 32.7MeV
\\
\Gamma (\Sigma^{\ast -} &\rightarrow &\Sigma ^{0}\pi ^{-})=\Gamma
(\Sigma^{\ast -}\rightarrow \Sigma^{\ast }\pi ^{0})\simeq
2.8MeV
\end{eqnarray*}%

Thus $\Gamma _{\Sigma^{\ast -}}\simeq 38.3MeV,$  to be compared
with the experimental value $\Gamma _{\Sigma^{\ast -}}\approx
(39.1\pm 2.1)MeV$  [3]. Finally, we get%
\begin{equation}
\frac{\Gamma (\Lambda _{b}\rightarrow \Delta ^{-}D^{+}\rightarrow n\pi
^{-}D^{+})}{\Gamma (\Lambda _{b}\rightarrow \Delta ^{0}D^{0}\rightarrow p\pi\Sigma
^{-}D^{0})}\approx 9  \label{47}
\end{equation}%

\begin{equation*}
\frac{\Gamma (\Lambda _{b}\rightarrow \Sigma^{\ast
-}D_{s}^{+}\rightarrow \Lambda \pi ^{-}D_{s}^{+})}{\Gamma (\Lambda
_{b}\rightarrow \Delta ^{0}D^{0}\rightarrow p\pi ^{-}D^{0})}\approx (0.28)%
\frac{(32.7)}{39}\approx 0.23  \label{48}
\end{equation*}%

For the second set of decays, we note that $\Xi _{c}^{0},\Xi
_{c}^{+},\Lambda _{c}$ belong to representation $\bar{3}$: and $\Sigma_{c}^{\ast
0},\Xi _{c}^{\ast 0},\Xi _{c}^{*\; +}$ belong to representation 6 of
SU(3). Now $\bar{3}\times 8\rightarrow 6+\bar{3}+15$, thus SU(3) gives%
\begin{equation*}
\Sigma_{c}^{\ast 0}\rightarrow \Lambda _{c}^{+}\pi ^{-}: -%
\sqrt{2}g_{c}^{\ast }
\end{equation*}
\begin{eqnarray*}
\Xi _{c}^{\ast 0\;, +}&\rightarrow& \Xi _{c}^{+\;, 0}\pi ^{-\;, +}: \pm g_{c}^{\ast } \\
&\rightarrow& \Xi _{c}^{0\;, +}\pi ^{0}:\mp \frac{1}{\sqrt{2}}g_{c}^{\ast }
\end{eqnarray*}%

First prediction of the above analysis taking into account phase space is
that total decay width of $\Xi^{*0}_{c}$:%
\begin{eqnarray*}
\Gamma _{\Xi _{c}^{\ast 0}} &\approx &\frac{3}{2}(0.11)\Gamma
_{\Sigma_{c}^{\ast 0}}=\frac{3}{2}(0.11)(14.5\pm 1.5)MeV  \notag\\
&=&(2.39\pm 0.25)MeV = \Gamma_{\Xi_{c}^{*+}}
\end{eqnarray*}%
{\large \newline
on using the experimental value }%
\begin{equation*}
\Gamma _{\Sigma_{c}^{\ast 0}}=\Gamma ({\large \Sigma_{c}^{\ast
0}\rightarrow \Lambda _{c}^{+}\pi ^{-}})=(14.5\pm 1.5)MeV
\end{equation*}%
{\large \newline
The experimental limits on decay width }$\Gamma _{\Xi _{c}^{\ast 0}}<5.5$%
{\large MeV, }$\Gamma _{\Xi _{c}^{\ast +}}< 3.1${\large MeV, [3]. 

Finally the branching ratios for three particle states $\Lambda _{c}^{+}\pi
^{-}\pi ^{0}$ and $\Xi _{c}^{+}\pi ^{-}K^{0}$ through resonances $%
\Sigma_{c}^{\ast 0}$ and $\Xi _{c}^{\ast 0}$ is given by \newline
}%
\begin{equation*}
\frac{\Gamma (\Lambda _{b}\rightarrow {\large \Xi _{c}^{\ast
0}K^{0}\rightarrow \Xi _{c}^{+}\pi ^{-}K^{0}})}{\Gamma (\Lambda
_{b}\rightarrow {\Sigma_{c}^{\ast 0}\pi ^{0}\rightarrow \Lambda
_{c}^{+}\pi ^{-}\pi ^{0}})}=(4.24)(0.11)\approx 0.47
\end{equation*}%

To summarize, we have analyzed two sets of decays of $\Lambda_b$:
\begin{equation*}
\Lambda_b \rightarrow \Sigma^{0}_c \rightarrow \Delta^0 D^0\; , \Delta^- D^+\;, \Sigma^{*-}D^{+}_s
\end{equation*}
\begin{equation*}
\Lambda_b \rightarrow \Sigma^{0}_c \rightarrow \Sigma^{*+}\pi^{-}\;, \Sigma_{c}^{*0}\pi^0\; , \Xi_{c}^{*0}K^0
\end{equation*}
We note that the other two members of triplet $\bar{3}$ are $\Xi_b^{0}=\frac{1}{\sqrt{2}}(us-su)b\chi_{MA}$ and  $\Xi_b^{-}=\frac{1}{\sqrt{2}}(ds-sd)b\chi_{MA}$. The $W-$exchange can not generate a baryon pole for $\Xi_b^{-}$, thus the decays $\Xi_b^{-}\rightarrow  B^*(\frac{3}{2}^+)+P$ are not possible. This is another prediction of our formalism. However, for $\Xi_{b}^0$, $W-$exchange give:
\begin{eqnarray*}
H_{W}^{p.c}|\Xi_{b}^{0}\rangle & = &-\frac{1}{2}[-2(d^\uparrow s^\uparrow +s^\uparrow d^\uparrow)c^\downarrow +(d^\uparrow s^\downarrow +d^\downarrow s^\uparrow )c^\uparrow + (s^\uparrow d^\uparrow +s^\downarrow d^\uparrow)c^\uparrow \notag \\
& = &\sqrt{6}|\Xi^{\prime 0}_{c}\rangle 
\end{eqnarray*}
Thus $\langle \Xi^{\prime 0}_{c}|H_{W}^{p.c.}|\Xi_b^{0}\rangle = \sqrt{6}$.

Hence the pole diagram gives two set of decays:
\begin{equation*}
\Xi_b^{0} \rightarrow \Xi^{\prime 0}_c \rightarrow \Sigma^{*0} D^0\; , \Sigma^{*-} D^{+}\;, \Xi^{*-}D_{s}^{+}
\end{equation*}
\begin{equation*}
\Xi^{0}_b \rightarrow  \Xi^{\prime 0}_c \rightarrow \Xi_c^{*+} \pi^-\; , \Xi_c^{*0} \pi^0\;, \Omega_c^{*0}K^0
\end{equation*}
Hence in SU(3) limit, for two sets of decays, $p-$wave (parity conserving) amplitude $C$ is given by
Set I: 
\begin{equation*}
C=\sqrt{2}(1,1,1)g_c\frac{<\Xi^{\prime 0}_{c}|H^{pc}_{W}|\Xi^0_{b}>}{%
m_{\Xi^0_{b}}-m_{\Xi^{\prime 0}_c}}
\end{equation*}
Set II: 
\begin{equation*}
C=\frac{1}{\sqrt{2}}(\sqrt{2},-1,2)g_c\frac{<\Xi^{\prime 0}_{c}|H^{pc}_{W}|\Xi^0_{b}>}{%
m_{\Xi^0_{b}}-m_{\Xi^{\prime 0}_c}}
\end{equation*}

Following exactly the same procedure as for the $\Lambda_b$ decays, one can calculate the branching ratios $\Xi^0_b \to B(\frac{3^+}{2}) + P$ decays. At present no experimental data for $\Lambda_b \to B^*(\frac{3^+}{2}) + P$ and $\Xi^0_b \to B^*(\frac{3^+}{2}) + P$ are available to test the prediction of our model. In future, it is expected that more data for heavy flavor hadron decays will be coming from LHCb including the decays considered in this paper.

{\bf{Acknowlegement}}

The author would like to thank Dr. Muhammad Jamil Aslam for discussions.

{\large \pagebreak }

\end{document}